\documentclass[twocolumn,preprintnumbers,amsmath,amssymb]{revtex4}
\input{epsf}

\newcommand{\be}{\begin{equation}}
\newcommand{\ee}{\end{equation}}
\newcommand{\bea}{\begin{eqnarray}}
\newcommand{\eea}{\end{eqnarray}}
\newcommand{\bef}{\begin{figure}}
\newcommand{\eef}{\end{figure}}

\newcommand{\simge}{\,{}^>_{\sim}\,}
\newcommand{\simle}{\,{}^<_{\sim}\,}

\def\be#1{$^{#1}$Be}

\def\eps@scaling{0.96}

\def\showone#1{
  \centering
  \leavevmode
  \epsfxsize=\eps@scaling\linewidth
  \epsfbox{#1.eps}
}

\def\epstwo@scaling{0.48}

\def\showtwo#1#2{
  \centering
  \leavevmode
  \epsfxsize=\epstwo@scaling\linewidth
  \epsfbox{#1.eps} \hfil
  \epsfxsize=\epstwo@scaling\linewidth
  \epsfbox{#2.eps}
}

\begin{document}

\title{Gravitino, Axino, Kaluza-Klein Graviton Warm and Mixed \\
Dark Matter and Reionization}

\author{{\sf Karsten Jedamzik} ${}^{a}$, {\sf Martin Lemoine} ${}^{b}$, 
{\sf Gilbert Moultaka} ${}^{a}$} 

\affiliation{
{\small {\it ${}^{a}$ Laboratoire de Physique Th\'eorique et Astroparticules,
CNRS UMR 5825,\\
Universit\'e Montpellier II, F-34095 Montpellier Cedex 5, France} }}
\affiliation{~~}

\affiliation{
{\small {\it ${}^{b}$ GReCO, Institut d'Astrophysique de Paris, CNRS, \\
98 bis boulevard Arago, F-75014 Paris, France}} }

\begin{abstract}
Stable particle dark matter may well originate during the decay of 
long-lived relic particles, as recently extensively examined in the
cases of the axino, gravitino, and higher-dimensional Kaluza-Klein (KK)
graviton. It is shown that in much of the viable parameter space
such dark matter emerges naturally warm/hot or mixed.  
In particular, decay produced gravitinos (KK-gravitons) may only be considered
cold for the mass of the decaying particle in the several TeV range,
unless the decaying particle and the dark matter particle 
are almost degenerate.
Such dark matter candidates are thus subject to a
host of cosmological constraints on warm and mixed dark matter, such
as limits from a proper reionization of the Universe, 
the Lyman-$\alpha$ forest, and the abundance of clusters of galaxies.. 
It is shown that constraints from an early reionsation epoch, such as
indicated by recent observations, may potentially limit such warm/hot
components to contribute only a very small fraction to the dark matter.

\end{abstract}


\maketitle

The nature of the ubiquitous dark matter is still unknown. 
Dark matter in form of fundamental, and as yet experimentally undiscovered, 
stable particles predicted to exist in extensions of the standard model of 
particle
physics may be particularly promising. For a scale of the new physics
around $1\,$TeV, as preferred by theoretical arguments, such dark matter
abundances produced either during freeze-out of stable particles
from equilibrium, 
or freeze-out of meta-stable particles and their subsequent decay 
into stable particles, may come tantalizingly
close to the abundance required by cosmology. For this reason, a number of
candidate dark matter particles, including also stable axinos produced 
in decays
of next-to-lightest supersymmetric particles (NLSP); binos or 
staus~\cite{Rajagopal:1990yx,Covi1,Covi:2001nw,Covi},
stable gravitinos produced via NLSP bino, stau, or sneutrino 
decays~\cite{Feng:2003xh,Feng:2003uy,Feng:2004zu,
Feng:2004mt,Roszkowski:2004jd}, or stable
KK-gravitons produced via the decay of KK $U(1)$ hypercharge gauge 
bosons~\cite{Feng:2003xh}, 
have been recently
considered/reconsidered. 
For details and the theoretical motivation for
the possible existence of such particles we refer the reader to the
orginal literature. 

In this note we consider the fact that particle dark matter (DM)
produced by the decay of a relic population of metastable particles is
often warm or even hot. This has been known
for some time~\cite{Borgani:1996ag}, but has escaped entering into the
conlusions of some recent studies. This effect is also known to exist
in the context
of dark matter production by cosmic string evaporation~\cite{Lin:2000qq}.
Nevertheless, very recently the
very same observation has been rediscovered in the studies by 
Cembranos {\it et al.}
~\cite{CFRT05} and Kaplinghat~\cite{K05}. These latter papers concentrate
on the possible resolution of a number of alleged deviations between the
observed sub-structure of galactic halos and structure of dwarf galaxies
and that predicted in structure formation with a purely cold DM particle,
whereas our study focusses mostly on constraining such warm- or mixed-
dark matter models by reionization. In any case, for a complete view the
reader is referred to the above studies as well.

Decay produced particle dark matter
is often warm/hot, i.e. is endowed with primordial free-streaming
velocities leading to the early erasure of small-scale 
perturbations~\cite{KT90}, due
to the kinetic energy imparted on the decay product during the decay
process itself. Since axinos, gravitinos, and KK-gravitons are superweakly
interacting, they will not
thermalise after decay, and inherit as kinetic energy a good fraction of
the rest mass energy of the mother particle. For decay 
$\chi\to\gamma + \tilde{G}$ of a massive particle $\chi$ to an essentially 
massless
particle $\gamma$ (with $m_{\gamma}\ll m_{\chi}$) 
and the dark matter particle $\tilde{G}$, with
$m_{\chi}>m_{\tilde{G}}$ one finds for the instantaneous post-decay momentum of
$\tilde{G}$ 
\begin{equation}
p_{\tilde{G}}^i = {{(m_{\chi}^2-m_{\tilde{G}}^2)}\over {2m_{\chi}}} \, .
\label{eq1}
\end{equation}
The momentum of particles of arbritrary relativity redshifts with
the scale factor of the Universe, and insisting for the particle to be
non-relativistic at the present epoch, one may compute the present
free-streaming velocity of the DM particle
\begin{equation}
v_{DM}^0 = \biggl({{{m_{\chi}^2-m_{\tilde{G}}^2}}\over 
{2m_{\chi}m_{\tilde{G}}}}\biggr)
\frac{T_0}{T_d}\biggl(\frac{g_0}{g_d}\biggr)^{1/3}\, ,
\label{Eq:v01}
\end{equation}
where $T$ and $g$ are cosmic temperarture and statistical weight of the
plasma at the present $'0'$ and particle decay $'d'$ epochs, respectively.
It has been shown that a sudden decay approximation, 
approximating all particles
to decay at the same $T_d$, is excellent~\cite{Covi:2001nw}, provided one 
uses the relation $t_d\simeq \Gamma_d^{-1}$ with $t = (2H)^{-1}$ in the 
early radiation dominated Universe, where $t$ is cosmic time, $\Gamma_d$ is
the decay width of the particle, and $H$ the Hubble constant, respectively.
Using the above one finds 
\begin{equation}
v_{DM}^0 \simeq 4.57\times 10^{-5}\, {\rm \frac{km}{s}}
\biggl({{{m_{\chi}^2-m_{\tilde{G}}^2}}\over 
{2m_{\chi}m_{\tilde{G}}}}\biggr)
g_d^{-1/12}
\biggl(\frac{\tau_d}{1\,\rm s}\biggr)^{1/2}
\,  .
\label{Eq:v02}
\end{equation}
Free-streaming velocities may become appreciable for either light DM
particles (i.e. $m_{\tilde{G}}\ll m_{\chi}$) or late decay, a condition
satisfied for much of the axino-, gravitino-, and KK-graviton- DM parameter
space.

Cosmological constraints on warm dark matter deriving, for example, from
the Lyman-$\alpha$ forest or cosmological reionization, are
often formulated as lower limits on the mass of a hypothetical gravitino
DM particle. Such limits assume a relativistically freezing 
out gravitino, leading thereby to a well-specified relation  
$v_{\rm rms}\simeq 0.044 {\rm km s^{-1}}(\Omega_{\tilde{G}} h^2/0.15)^{0.34}
(m_{\tilde{G}}/{1\, keV})^{-1.34}$~\cite{remark} between gravitino mass 
and root-mean-square present day gravitino 
velocity (cf.e.g. to~\cite{Bode:2000gq,Barkana:2001gr}). However, it is
rather the latter quantity, $v_{\rm rms}$, which is, without further
assumptions about the nature and production mechanism of the dark matter,
constrained by cosmology. Free-streaming particles erase
primordial dark matter perturbations between very early times and the epoch
of matter-radiation equality (EQ), after which further erasure becomes
inefficient. Up to which scale perturbations have been erased than depends
essentially only on $v_{\rm rms}$ and the 
time of EQ (determined itself by $\Omega_{m}$,
the total non-relativistic matter density), such that one 
finds~\cite{Barkana:2001gr}
\begin{equation}
R_c^0 \simeq 0.226 {\rm Mpc}\,\biggl(\frac{\Omega_m}{0.15}\biggr)^{-0.14}
\biggl(\frac{v_{\rm rms}}{0.05\,{\rm km/s}}\biggr)^{0.86}\, ,
\label{Eq:Rc0} 
\end{equation}
by detailed Boltzman-equation 
simulations~\cite{Ma:1995ey}. Here $R_c^0\equiv 1/k_c$ defines
the wave vector $k_c$ for which the primordial power spectrum is suppressed
by a factor two~\cite{remark2} 
when compared to the same cosmological model, but with cold
dark matter. 
Armed with a relation between $v_{\rm rms}$ and gravitino mass, as well
as Eq.~(\ref{Eq:v02}), we may now translate in the literature existing
limits on relativistically freezing out gravitino warm dark matter, into
limits on warm dark matter generated by particle decay. Here we employ
decay widths as calculated in the original literature~\cite{remark3}.

We note here that the equivalence between traditional warm dark matter,
described by a Fermi-Dirac distribution, and metastable particle decay
produced dark matter, with a velocity distribution given by exponential
decay, is not entirely perfect. This is due to the differing velocity
distributions. Kaplinghat~\cite{K05} computes the tansfer function for
decay produced dark matter and finds differences in the damping tails
between decay produced dark matter and warm dark matter. Nevertheless,
given current measurement and theoretical uncertainties in cosmology,
such as for example in the determination of the epoch of reionization,
both types of DM should be considered as having equivalent effects on
structure formation as long as they have an identical second moment
of the distribution, i.e. an identical root-mean-square velocity. 

There exists a large number of observable cosmological differences between
scenarios with warm- (WDM) and cold- dark matter (CDM). 
It has even been argued
that WDM has phenomelogical advantages over CDM, potentially resolving
possible difficulties of CDM scenarios in explaining a scarce of substructure
in Milky-Way type halos or the existence of cores in 
dwarf galaxies.
Nevertheless, counterarguments in favor of CDM have also 
been presented, 
such that the situation is not resolved. We will here only focus on three
cosmological implications of WDM or mixed dark matter (MDM) scenarios; 
namely the
optical depth in the Lyman-$\alpha$ forest of mildly non-linear fluctuations
on smaller scales $\sim 1\,$Mpc~\cite{Viel:2005qj}, the successful
reionization of the Universe at high redshift (probing perturbations
on the 
smallest scales $\sim 10-100\,$kpc)~\cite{Barkana:2001gr,Yoshida:2003rm}, 
and for the case of MDM, the
abundance of clusters at close to the present epoch~\cite{KTAR05}. 
A recent analysis
of the matter power spectrum as implied by observations of
the Lyman-$\alpha$ forest and the cosmic microwave background anisotropies
(CMBR) by the WMAP mission has yielded a $2\sigma$ lower limit on the
WDM gravitino mass of $550\,eV$~\cite{Viel:2005qj}. 
Using the above this may be translated to
a limit $v_{\rm rms}^0 \simle 0.1\,{\rm km\, s^{-1}}$. 
One year of observations of polarization
of the CMBR by the WMAP sattelite have revealed a high optical depth 
$\tau\approx 0.17\pm 0.04$~\cite{Kogut:2003et} 
for Thomson scatterings of CMBR photons on electrons.
The recently presented three year WMAP analysis has led to a downward revision
of this value to $\tau\approx 0.09\pm 0.03$~\cite{Spergel}, where error bars
are one sigma and a $\Lambda$CDM concordance model has been assumed. 
This implies a fairly complete reionization of the Universe at redshift
$8.5\simle z\simle 15$ (at $95\%$ confidence level) 
seemingly consistent with CDM scenarios. Such scenarios 
predict an early reionization of the Universe due to the early formation of 
sub-galactic halos and massive stars 
therein. The situation is different in WDM scenarios due to the lack of 
small-scale power and the concomitant late formation of the first stars.
If the fairly 
high optical depth is indeed due to the first stars, rather than due
to some 'exotic' mechanism, reionization places very stringent limits
on the warmness of the DM. In particular, Yoshida 
{\it et al.}~\cite{Yoshida:2003rm} have shown that
even for WDM with $m_{\tilde{G}}\approx 10\,$keV reionization 
is far from substantial at redshift $z\sim 17$ (appropriate to the
$\tau$ central value of the one year WMAP analysis). This corresponds to
$v_{\rm rms}^0 \simle 0.002\,{\rm km\, s^{-1}}$. In the numerically expansive
study of Yoshida et al WDM scenarios with $m_{\tilde{G}} > 10\,$keV
have not been considered. As the case $m_{\tilde{G}} = 10\,$keV fails,
even larger $m_{\tilde{G}}$ could potentially fail. Nevertheless, uncertainties
in these calculations exist due to the modeling of the 
physics of gas cooling, radiation
transport, and star formation. We thus regard such limits as preliminary.

\bef
\epsfxsize=9cm
\epsffile[50 50 410 302]{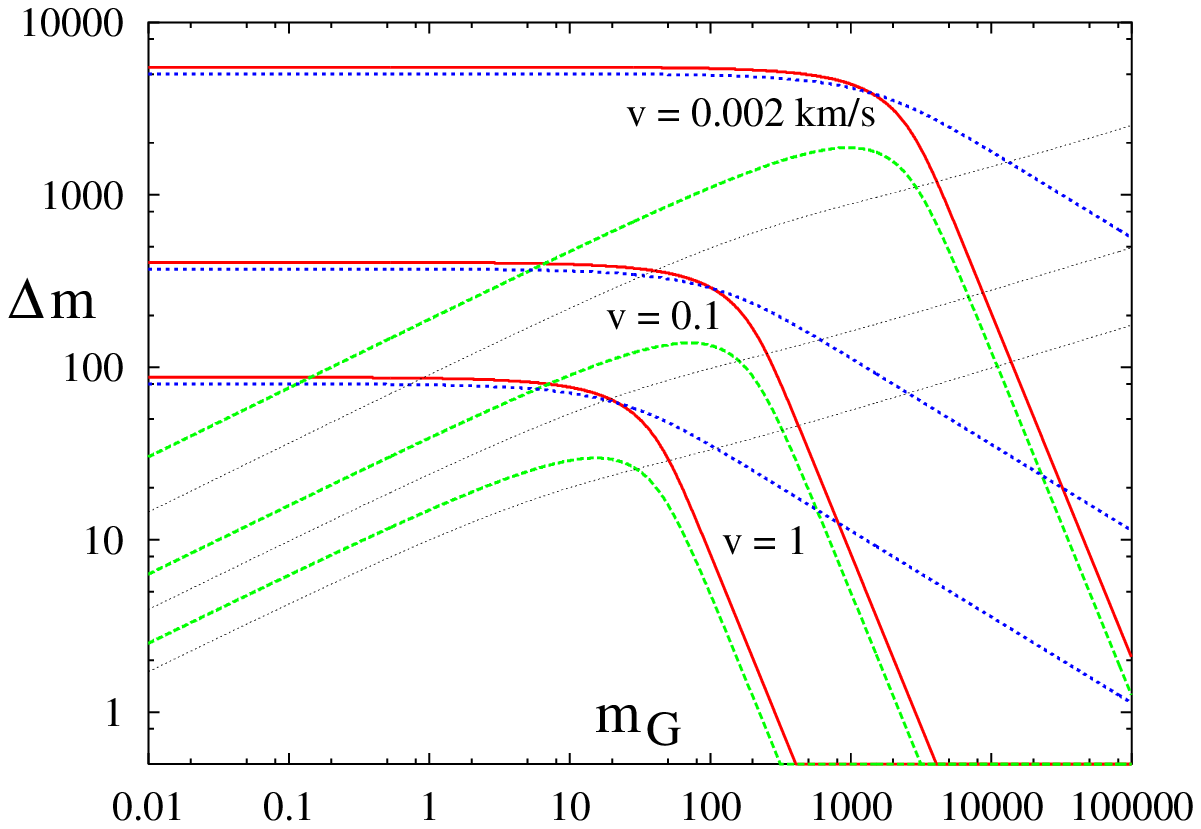}
\caption{Contour plots of constant present-day free-streaming velocities
in a variety of scenarios where the dark matter $\tilde{G}$
is generated by the late decay of a non-relativistic 
primary $\chi\to\tilde{G} + \gamma$ (cf.Eq.\ref{eq1}). Results are shown
in the plane of $m_{\tilde{G}}$ and $\Delta m\equiv m_{\chi}-m_{\tilde{G}}$
(all in GeV). The scenarios are bino-decay into gravitinos (red-solid),
slepton decay into gravitinos (blue-dotted), and $B^1$ decays into
KK-gravitons (green-dashed). Shown are the contours of velocities
(from top to bottom) $v = 0.002,0.1,$ and $1\,$ km s$^{-1}$, respectively.
Also shown, by the thin dotted lines, are the contours where the effects 
of small-scale suppression due
to the coupling of a charged slepton to the CMBR decaying later
into a gravitino
are similar to warm dark matter with 
$v = 0.002,0.1,$ and $1\,$ km s$^{-1}$, respectively 
(see text for details).}
\label{fig1}
\eef

Scenarios of DM due to decaying metastable
particles often may come in the flavor of mixed dark matter when only a
component of the DM is due to decay, with the other component possibly due
to thermal scatterings at high temperature.
Such scenarios of MDM scenarios may also be constrained by the abundance
of clusters of galaxies~\cite{KTAR05}. 
Nevertheless, MDM may only be constrained by these
means, in case the warm component is warm/hot enough to erase primordial
perturbations on the scale of a cluster of galaxies. For 
$\lambda_c\equiv 2\pi/k\approx 20\,$Mpc we find that a half wavelength
roughly encompasses $10^{14}M_{\odot}$, the approximate mass of a typical
cluster. In order to erase perturbations on the scale $\lambda_c$ the present
day DM velocity has to exceed $v_{\rm rms}^0 \simge 1\,{\rm km\, s^{-1}}$.
If this is the case, however, only a fraction between $10-20\%$ of all the
DM may be warm/hot~\cite{KTAR05} (cf. also to~\cite{Viel:2005qj} for
similar limits from the Lyman-$\alpha$ forest).

In Fig. 1 WDM limits on gravitino DM produced by metastable particle decay
are shown in the parameter space 
spanned by the mass splitting between the gravitino and the NLSP
$\Delta m\equiv m_{\rm NLSP} - m_{\tilde{G}}$ and the gravitino mass itself.
The respective limits of $v_{\rm rms}^0 = 0.002,\, 0.1,\, 
{\rm and}\, 1
{\rm km s^{-1}}$, as discussed above, 
are indicated as heavy lines for the cases when either the
bino (long-dashed) or the stau or sneutrino (solid) is the NLSP. The order
of the lines is such that $v_{\rm rms}^0$ exceeds the limit in the parameter
space below the lines, with lines at higher $\Delta m$ 
corresponding to a limit with
lower $v_{\rm rms}^0$. It is seen that most of the parameter space results
in WDM, potentially already in conflict with, at least, the high optical
depth as inferred by WMAP. In particular, only for excessively large 
$m_{\rm NLSP} \simge 5\,$TeV decay produced gravitino dark matter may be
considered almost cold. 
The limit may be less stringent when only a small component
$\simle 10\%$ of the gravitino DM is due to decay, as then masses 
$m_{\rm NLSP}\simge 100\,$GeV ascertain that the decay-produced component
is not too hot to delay the formation of clusters of galaxies.

We note here, however, that for gravitino masses not too small 
($m_{\tilde{G}}\simle 0.1\,$ GeV for a bino NLSP and 
$m_{\tilde{G}}\simle 10\,$ GeV for a stau NLSP~\cite{Feng:2004zu,Feng:2004mt}) 
very stringent constraints on such scenarios apply from a disruption of Big
Bang nucleosynthesis (BBN) with smaller mass splittings $\Delta m$
preferentially constrained by the effects of electromagnetic energy
injection after the epoch of BBN and distortion of the CMBR blackbody
spectrum and larger $\Delta m$ by hadronic three-body decays during and after
BBN. In particular, 
$^6$Li~\cite{BBN6Li} production as well as significant perturbations 
of the $^3$He/D~\cite{BBN3HeD} 
ratio potentially rule out much of the parameter space
at larger $m_{\tilde G}$. If such constraints are combined with the
requirement of having the supersymmetric potential bounded from below,
only very little parameter space remains, at least in the constrained
minimal supersymmetric standard model (CMSSM) and for small 
tri-linear couplings $A$~\cite{Cerdeno}.
These limits are not shown in Fig. 1.

Fig.1 shows also analogous results for a KK-$B^1$ decaying into a photon 
and KK-graviton, with constraints indicated by the dotted line. It is seen that
constraints from the warmness are not quite as stringent as in the gravitino
case, yet, for $m_{G^1}$ around the weak scale, $m_{B^1}\simge 1\,$TeV is
still required for a sucessful reionization by star formation. Only for
much smaller $m_{G^1}\simle 1\,$GeV may decay of lighter 
$m_{B^1}\simle 300\,$GeV result in CDM. 
However, it is rather expected for
the mass difference between the KK-$B^1$ and KK-graviton to be small, since
it should be only due to radiative corrections.  
Almost degenerate $B^1$ and $G^1$ are then constrained by limits from BBN.
It has recently been pointed out that a generation of dark matter by
particle decay may lead to an additional suppression of small-scale power
provided the decaying particle is charged~\cite{Sigurdson:2003vy}. 
Since the charged
'primary' is coupled to the CMBR sub-horizon perturbations
in the primary-photon fluid are below the Jeans mass. 
Perturbations may thus only start growing by the
gravitational instability when decoupled from the CMBR, i.e. after the
decay of the primary. 
In particular, it was found in Ref.~\cite{Sigurdson:2003vy,Profumo:2004qt} 
that for a charged particle decay time 
of $\tau_d = 3.5\,$yr the wavevector $k_d$ where the power
is reduced by half, when compared to a CDM scenario, is approximately
$k_d \approx 3\,$Mpc$^{-1}$ (assuming $h=0.7$). 
Comparing this to Eq.(\ref{Eq:Rc0}) one may
infer a velocity $v_{\rm rms}^0\approx 0.078\,{\rm km\, s^{-1}}$ which would
yield a similar erasure of small-scale power
due to finite DM velocities. An analogy between the net result of
these two physically different small-scale power suppression mechanism may
be established by noting that the
damping scale due to charged particle decay is approximately the
horizon scale at decay~\cite{Sigurdson:2003vy}, $\lambda_d\sim 0.265 {\rm Mpc}
(\tau /{\rm yrs})^{1/2}$.
Comparing this scale to Eq.~(\ref{Eq:Rc0})
the above limits of $v_{\rm rms}^0 = 0.002,\, 0.1,\, 
{\rm and}\, 1 {\rm km s^{-1}}$ translate to decay times
$\tau_d\sim 3\times 10^{-3},\, 2,\,{\rm and}\, 125\,$yr, respectively.
In Fig.1 the light lines are lines of constant decay
time for the decay $\tilde{\tau}\to \tau + \tilde{G}$ 
with values as given above. Here shorter decay times are at higher $\Delta m$. 
It is seen that in the scenarios considered here the effects of free-streaming
are generally more important than those of charged particle decay, though
this may be different when other scenarios are 
considered~\cite{Profumo:2004qt}.

We now investigate the resulting free-streaming velocities in
axino DM generation due to stau- ($\tilde{\tau}\to\tau + \tilde{a}$) and bino- 
($\tilde{B}\to\gamma + \tilde{a}$) decays as proposed by
Ref.~\cite{Rajagopal:1990yx,Covi1,Covi:2001nw,Covi}.
Taking the decay widths of the literature~\cite{remark5} we find
present root-mean-square velocities of
\bea
v_0\approx 14.7 {\rm \frac{km}{s}}
\biggl(\frac{m_{\tilde{a}}}{1\, {\rm MeV}}\biggr)^{-1}
\biggl(\frac{m_{\tilde{\tau}}}{100\, {\rm GeV}}\biggr)^{1/2} 
\quad\quad\quad\nonumber \\
\times \biggl(\frac{m_{\tilde{B}}}{100\, {\rm GeV}}\biggr)^{-1} 
\biggl(\frac{f_{{a}}}{10^{11} {\rm GeV}}\biggr) \label{Eq5}\\
v_0\approx 1.68 {\rm \frac{km}{s}}
\biggl(\frac{m_{\tilde{a}}}{1\, {\rm MeV}}\biggr)^{-1}
\biggl(\frac{m_{\tilde{B}}}{100\, {\rm GeV}}\biggr)^{-1} 
\biggl(\frac{f_{{a}}}{10^{11} {\rm GeV}}\biggr) \label{Eq6}
\eea
for stau- Eq.~(\ref{Eq5}) and bino- Eq.~(\ref{Eq6}) decay, respectively.
Here $m$ denote the masses for axino $\tilde{a}$, stau $\tilde{\tau}$ and
bino $\tilde{B}$, and $f_a$, required to be in the range 
$10^{10}{\rm GeV}\simle\, f_a \simle\, 10^{12}{\rm GeV}$, is the scale
where the chiral $U(1)$-symmetry, related to the Peccei-Quinn symmetry is
spontaneously broken. In the above we have neglected numbers of order
unity and a very weak dependence on the statistical weight during decay.
Velocities become excessively large for small $m_{\tilde{a}}$. 
It is evident that axinos generated by decay
having masses $m_{\tilde{a}}\simle 10\, (1)\,$GeV should be considered as warm
and may be in conflict with an early reionization by stars. However,
since the decay contribution to the axino density 
$\Omega_{\tilde{a}}^{decay} =  \Omega_{\tilde{B},\tilde{\tau}}
m_{\tilde{a}}/m_{\tilde{B},\tilde{\tau}}$ becomes typically small when
$m_{\tilde{a}}$ is small and moreover efficient production of ``thermal''
axinos during reheating is possible at low temperatures for small 
$m_{\tilde{a}}$~\cite{Brandenburg:2004du}, axino dark matter should be naturally
mixed, rather than warm,
with a small component of axinos with large velocities, and the rest
essentially cold. 
Note that due to the relatively large decay width,
suppression of small-scale power due to existence of a 
charged particle primary~\cite{Sigurdson:2003vy} is generally
unimportant for axinos. 

We have observed that constraints on warm dark matter
from cosmic reionization may become very stringent, particularly if an
early reionization epoch as indicated by the excessive low multipole
CMBR polarization in the WMAP data~\cite{Kogut:2003et} is confirmed.
On the other hand, gravitino- and axino- dark matter scenarios often
are coming as mixed dark matter scenarios with only a smaller fraction
of the dark matter warm/hot. As we will see, even in this case future
constraints could be very stringent.

\bef
\epsfxsize=9cm
\epsffile[54 360 558 760]{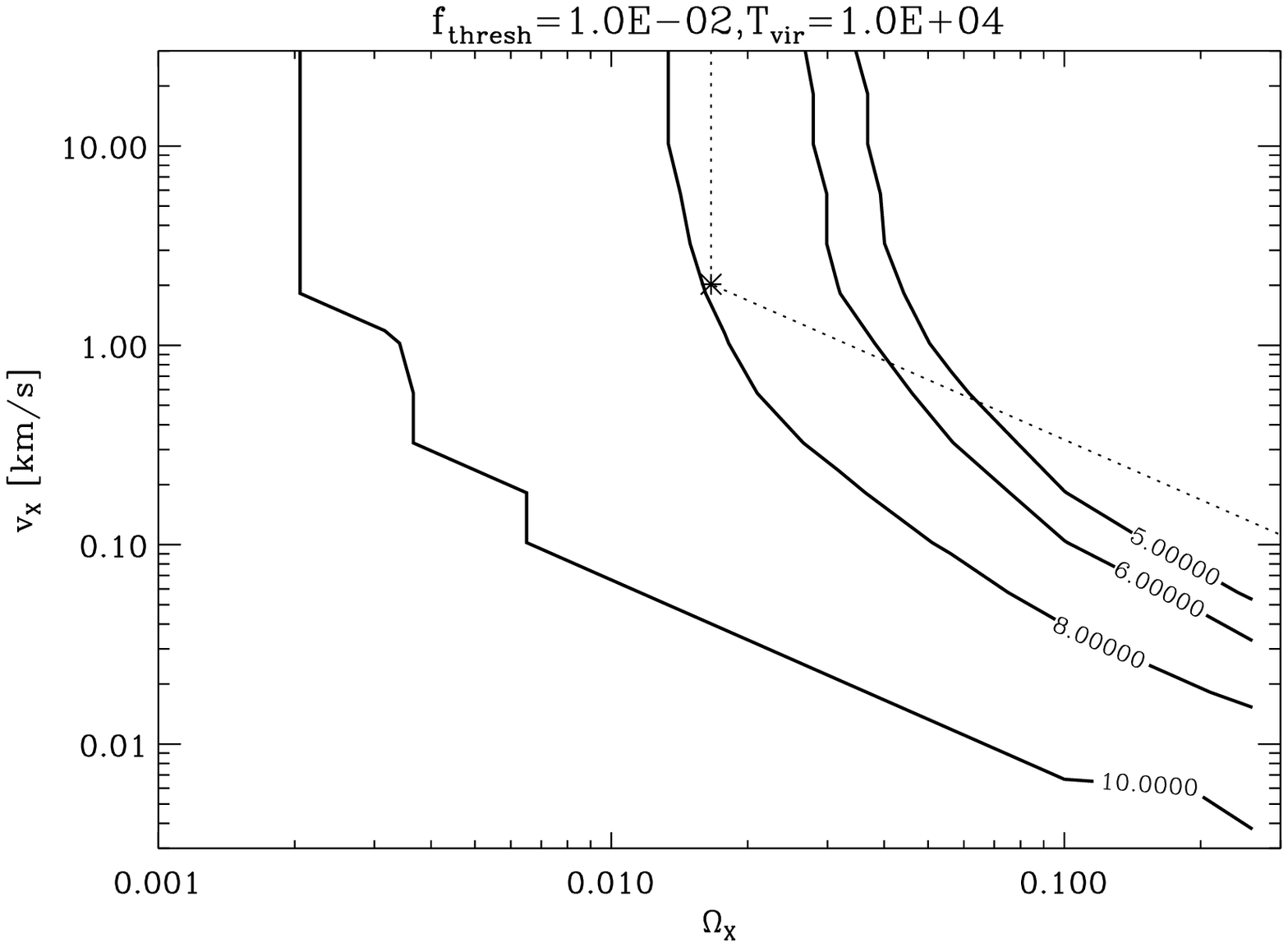}
\caption{Contour plot of the redshift $z$ when a fraction $10^{-2}$ of all
baryons is collapsed within halos exceeding a virial temperature of
$10^4$K. Results are shown depending on 
the warm/hot dark matter fractional contribution to
the present critical density $\Omega_x$ and the present free-streaming velocity
$v_x$ in km s$^{-1}$ of the warm/hot component. 
The calculation assumes the remainder of the dark matter
$\Omega_c = 0.26 - \Omega_x$ to be cold. The shown 
redshifts may be identified with
the approximate reionsation redshift when a ``standard'' reionsation efficiency
is assumed. For comparison, the WMAPIII results of optical depth
$\tau = 0.09\pm 0.03$ imply a reionization redshift within the $\Lambda$CDM
concordence model between $z\approx 8.5-15$ at 95\% confidence level.
The star indicates a recent limit~\cite{Viel:2005qj}
on a small contribution of $m_{\tilde{G}}\simge 16\,$ eV 
thermal gravitinos to the dark 
matter ruled out by a combination of CMBR and Lyman-$\alpha$ forest data. 
Regions above and right of the dashed line should be ruled out by these
considerations (see text).} 
\label{fig2}
\eef

\bef
\epsfxsize=9cm
\epsffile[54 360 558 760]{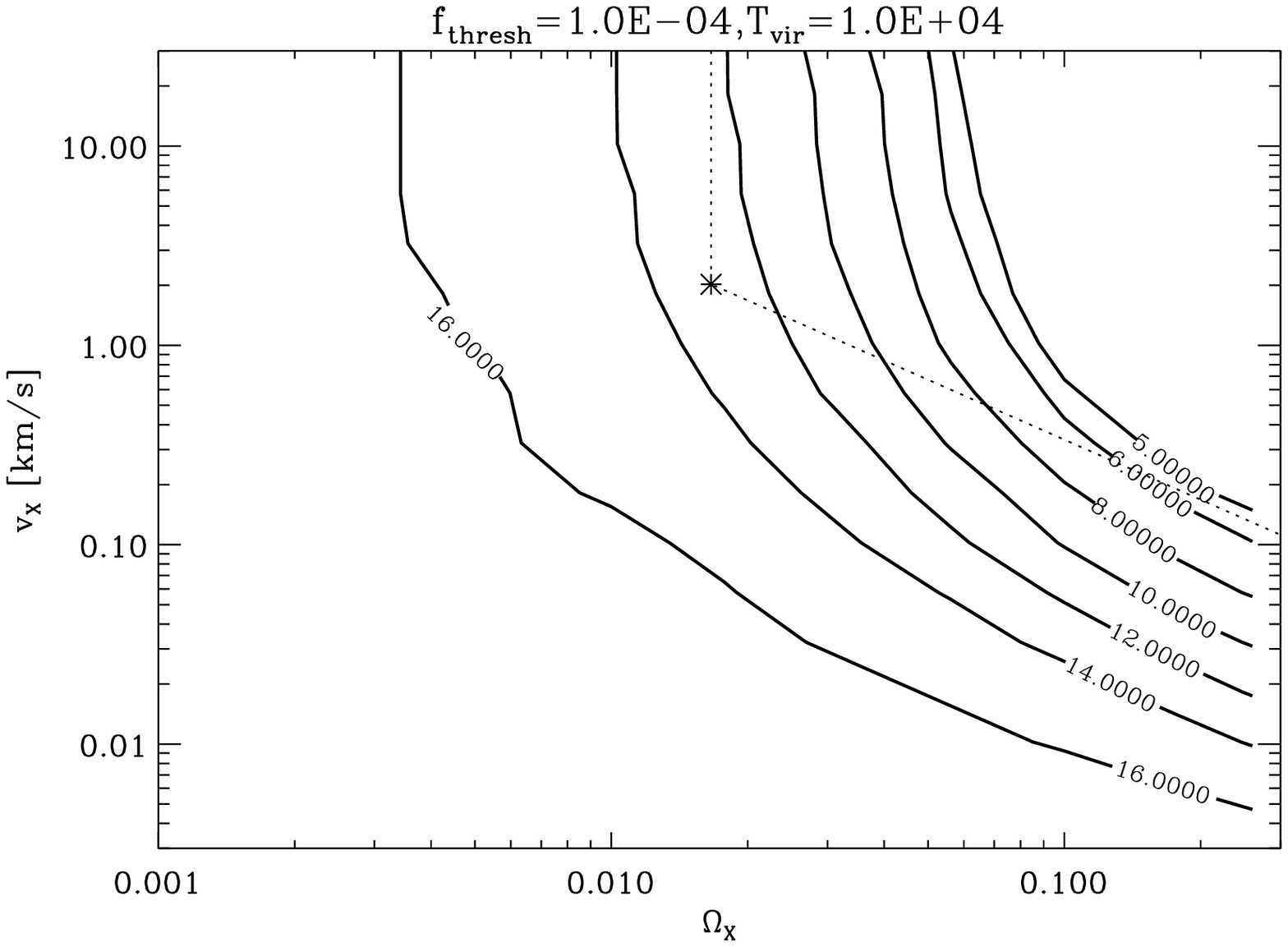}
\caption{As Fig.2, but for a collapse fraction $10^{-4}$,
corresponding to a factor $\sim 100$ more efficient reionization than inferred
from present-day properties of star clusters and star formation.} 
\label{fig3}
\eef

The Universe is believed to have been reionised mostly due to UV-radiation
by the first stars, with likely only a small contribution due to the
UV-radiation emitted by quasars~\cite{Barkana:2001gr}. 
It has been shown~\cite{HAR01} that the collapse of the smallest halos and
the stars which form therein actually impede further star formation as
the principal cooling agent $H_2$ is photodissociated. Only when harder
radiation $\sim 1\,$keV is emitted at the same time, 
cooling of baryons in small halos is still possible.
Efficient star formation and reionization thus generally may only occur
when larger halos, with virial temperatures $T_{vir}\simge 10^4$K, may
collapse as in such halos cooling is possible due to atomic hydrogen.
The typical UV-radiation production for a star 
cluster at low metallicity and for an initial mass function close to 
that observed locally is estimated around $N_{\gamma} = 4000$ 
ionising photons per stellar 
proton. Since normally only a small fraction $f_{\star}\sim 10\%$
of the gas forms stars and moreover only a small fraction $f_{esc}\sim 10\%$ is
capable of leaving the host galaxy, one may estimate around 
40 ionising photons per collapsed and cooled baryon. This implies
that the fraction of halos $F(T_{vir}> 10^4)$ with virial temperature larger
than $10^4$K has to be rather large, 
$\simge\, 1/(N_{\gamma}f_{\star}f_{esc}) = 2.5\times 10^{-2}$. Note that
this conservatively neglects further recombinations. A virial temperature
of $10^4$K corresponds to a mass scale of
$M_{10^4} \approx 3\times 10^7 M_{\odot}[(1+z)/11]^{-3/2}$ where
$z$ is redshift.

\bef
\epsfxsize=9cm
\epsffile[54 360 558 760]{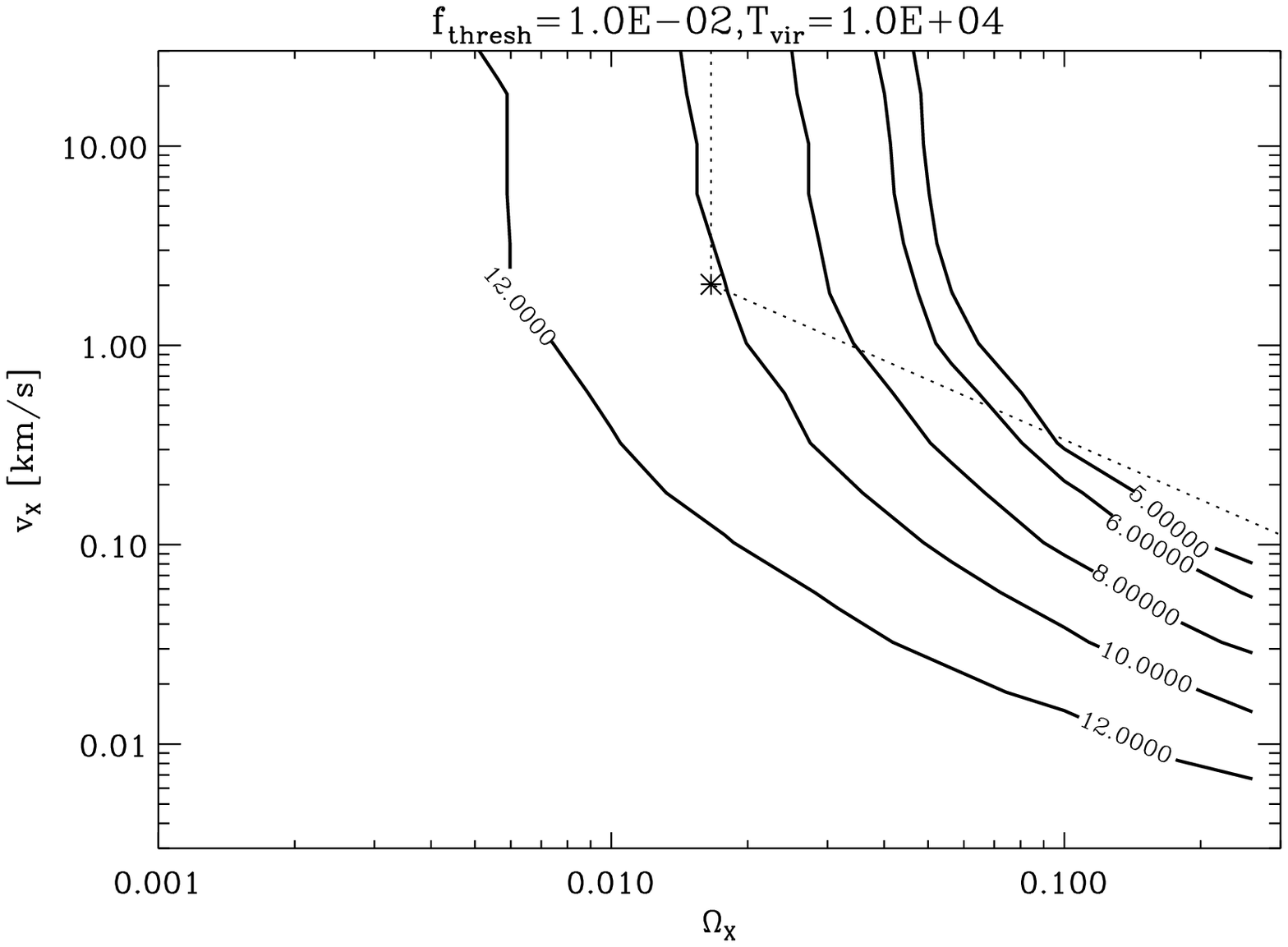}
\caption{As Fig.2, but for a spectral index for the adiabatic
primordial perturbations of $n_s=1$.} 
\label{fig4}
\eef

We have utilised CMBFAST in order to find the transfer function in models of
mixed dark matter with varying warm/hot dark matter fractions
$\Omega_x$ and root-mean-square velocity $v_x$. The transfer 
function allowed us
to compute the collapsed mass fraction $F(M)$ as a function of redshift.
Our models assume $\Omega_x + \Omega_c +\Omega_b = 0.3$, where 
$\Omega_c$  is the cold dark matter density parameter and $\Omega_b$ is the
baryonic density parameter, and we assumed $\Omega_b=0.04$ and a Hubble
parameter of $h=0.65$. Most of our calculations assume a red spectral
index for scalar adiabatic perturbations $n_s = 0.951$ corresponding to
the central value of the by the after three years of WMAP observations
determined $n_s = 0.951^{+0.015}_{-0.019}$
within the $\Lambda$CDM concordance model. This model assumes a negligible
contribution of primordial gravitational waves to the CMBR anisotropies.
Gravitational waves due to an inflationary epoch mostly contribute to the
CMBR anisotropies on large scales. If present, they may therfore also
describe the WMAP data well, albeit with a lower normalisation of scalar
perturbations on the largest scales and a larger spectral index $n_s$.
As the two sigma upper limit on $n_s$ from a combination of WMAP observations
and large-scale structure surveys 
(and in the absence of a running spectral index) is close to one, we have
also utilised a Harrison-Zeldovich spectrum $n_s = 1$ for one particular
computation. This may be indicative of variations of our results with
respect to the assumed cosmological model. All models are normalized
on the present horizon scale (taking COBE-DMR results, i.e. Eq.(30)
in Ref.~\cite{BW97}), irrespective of if gravitational waves are present or 
not. 
Finally, the warm component was simulated by adding a
massive ``neutrino'' to the fluid with a
root-mean-square velocity of $v_x$, while keeping also 
the three massless standard model neutrinos.

In Fig. 2 we show the redshifts at which in a particular mixed model 
(defined by $\Omega_x$ and $v_x$) a
fraction $> 10^{-2}$ of all baryons have collapsed within halos
exceeding a virial temperature of $10^4$K. Assuming an essentially
instantaneous baryon cooling and star formation~\cite{remark6}, 
and given the required 
$F(T_{vir}> 10^4)> 2.5\times 10^{-2}$ as inferred in a ``standard'' 
reionization scenario, this redshift $z$ may be approximately indentified
with the epoch of reionization. It is seen that even for warm/hot dark matter
fractions as small as $\Omega_x\sim 10^{-3}-10^{-2}$ reionization with
standard efficiencies may be problematic for redshifts larger than $z> 10$.
This should be compared to the current estimate of $z_{reion}\approx 8.5-15$ as
inferred by WMAP, indicating the potentially stringent nature of limits one
could potentially  
derive on very small warm/hot dark matter fractions from reionization.
Such limits, if holding up againts future observational tests, could
severely constrain the nature of particle dark matter, by requiring it to
be essentially exclusively cold. 

However, it needs to be stressed here, 
that the results as shown in Fig. 2 (as well as in
the figures which follow) are not intended by the authors to be employed
to derive limits on mixed dark matter models. This is due to a variety of
reasons, such as uncertainties in the reionization process and cosmological
model (see below), but also due to the fact that we have not marginalized over
all cosmological parameters entering the analysis. Furthermore, different
constraints, not shown in the figures, may be of importance. For example,
models with relatively large $v_X$ and $\Omega_X$, corresponding to a
significant hot dark matter component, may potentially be ruled out simply
by a relative mismatch between the predicted and observed powers on  
$8\,{\rm Mpc}\,h^{-1}$ and the present horizon scale $3000\,{\rm Mpc}\,h^{-1}$,
respectively. This is due to the hot component erasing power on 
$8\,{\rm Mpc}\,h^{-1}$ due to free-streaming.

Intrinsic uncertainties about
the reionization process during the dark ages are still large enough to
evade drastic conclusions. It may be that either the effiency of star
forming regions to produce UV-radiation (by a top-heavy initial mass
function, for example) or the escape fraction/star formation effiency are
substantially larger at early times than at the current epoch. Most of
such evolutionary effects should be testable, by, for example searches for
very high redshift galaxies or heavy element abundances in the high redshift
gas, by the NGST (next-generation space telescope). In Fig. 3 we show the
redshifts where a much smaller fraction of $10^{-4}$ 
of the gas was able
to collapse in virial halos with $T_{vir}> 10^4$K. This corresponds to a factor
$100$ increase in ``reionization'' efficiency. It is seen that fairly early
$z\sim 16$ reionsation epochs are possible except when $\Omega_x$ becomes
large $\simge\, 10^{-2}$ and free-streaming velocities are not too small 
$v_x \simge\, 0.1\,$km s$^{-1}$. 
It is evident, that even with a factor $\sim 100$ higher 
efficiency reionization than expected, and
provided the reionization redshift is found to 
be larger than $z\simge\, 12$, limits from reionization on warm/mixed 
dark matter may be more stringent than those derived from the 
CMBR-Lyman-$\alpha$ forest. This conclusion is also in accord with the
numerical simulations recently performed in Ref.~\cite{Yoshida:2003rm}.

Both plots show by the star and the lines
also the recently derived limit~\cite{Viel:2005qj} 
on a small thermal warm/hot gravitino component
from a combination of WMAP-CMBR data (probing perturbations on larger scales
$\sim 10-3000\,$Mpc)
and the Lyman-$\alpha$ forest (probing perturbations on 
smaller scales $\sim 1-40\,$Mpc). 
The models considered in Ref.~\cite{Viel:2005qj}
are lying on the diagonal dotted line and are ruled out because of a too
severe suppression of the power spectrum at small scales, at odds with
observations of the Lyman-$\alpha$ forest. Models with the
same $\Omega_X$ but even larger $v_X$ than that indicated by the star, 
moving the power spectrum supression to even larger scales, 
should also be at odds with the observations. We have therefore added a
vertical dotted line, such that
the region in the right upper hand corner
of the star (bounded by these dotted lines) should be ruled out.

Similarly, uncertainties in the prediction of the reionization redshift
in mixed dark matter models are also due to uncertainties in the cosmological
model and/or scalar spectral index $n_s$. In Fig. 4 we show the reionization
redshift in our simple reionization model for the same required collapse
fraction and virial halo temperatureas as in Fig. 2, but for a model which
has a Harrison-Zeldovich $n_s$ spectral index. It is seen that the
requirement of early reionization is more easily satisfied for $n_s=1$ than
for $n_s=0.951$. Nevertheless, the change is not big and amounts only to
$\Delta z_{reion} \approx 2$. We have also varied the COBE normalization,
moving it approximately two sigma upwards (resulting in a factor 1.14
larger perturbations) and have obtained very similar results to those shown
in Fig. 4, i.e. approximately $\Delta z_{reion} \approx 2$. We conclude
that the proper reionization of the Universe seems a promising alley
in constraining warm and mixed dark matter scenarios.
Finally we note that the potentially strong constraints on warm/mixed
dark matter possible from a determination of the reionsation redshift
have also been noted in the two very recent papers of 
Refs.~\cite{CFRT05,K05}



\begin{thebibliography}{99}
\bibitem{remark} Here $\Omega_{\tilde{G}}$ is the gravitino contribution
to the present critical density and $h$ is the Hubble constant
in units of 100 km s$^{-1}$ Mpc$^{-1}$.
\bibitem{remark2} The linear power spectrum for warm dark matter (WDM)
is given by the power spectrum for cold dark matter (CDM) multiplied
by a transfer function 
$T_{\rm WDM} = (1+(\epsilon k R_c^0)^{2\nu})^{-\eta/\nu}$ which accounts
for the additional free-streaming in WDM scenarios. In the above, $k$ is
wave vector and the values $\eta = 5$, $\nu = 1.2$, and $\epsilon = 0.361$
are frequently used.
\end{thebibliography}
\end{document}